\documentclass[preprintnumbers,prl,twocolumn,showpacs,amsmath,amssymb,superscriptaddress]{revtex4}
\usepackage{bm}
\usepackage{graphicx}
\usepackage{hyperref}

\newcommand{\rmd}{\mathrm{d}}   
\newcommand{\rmi}{\mathrm{i}}   
\newcommand{\GF}{G_{\mathrm{F}}} 
\newcommand{\dmsol}{\Delta m_\odot^2} 
\newcommand{\dmatm}{\Delta m_\mathrm{atm}^2} 

\newcommand{\myfigsep}{0.02 \textwidth}
\newcommand{\myfigwid}{0.38 \textwidth}

\begin{document}
\preprint{INT PUB 07-23}
\preprint{LA-UR-07-6726}

\title{Flavor Evolution of the Neutronization Neutrino Burst from
an O-Ne-Mg Core-Collapse Supernova}
\newcommand*{\INT}{Institute for Nuclear Theory, %
University of Washington, %
Seattle, WA 98195}
\affiliation{\INT}
\newcommand*{\UCSD}{Department of Physics, %
University of California, San Diego, %
La Jolla, CA 92093}
\affiliation{\UCSD}
\newcommand*{\LANL}{Theoretical Division, Los Alamos National Laboratory, %
Los Alamos, NM 87545}
\affiliation{\LANL}
\newcommand*{\UMN}{School of Physics and Astronomy, %
University of Minnesota, Minneapolis, MN 55455}
\affiliation{\UMN}

\author{Huaiyu Duan}
\affiliation{\INT}
\author{George M.~Fuller}
\affiliation{\UCSD}
\affiliation{\INT}
\author{J.~Carlson}
\affiliation{\LANL}
\author{Yong-Zhong Qian}
\affiliation{\UMN}


\begin{abstract}
We present results of 
3-neutrino flavor evolution simulations for the neutronization burst
from an O-Ne-Mg core-collapse supernova.
We find that nonlinear neutrino self-coupling engineers a single 
spectral feature of stepwise conversion
in the inverted neutrino mass hierarchy case and in the normal
mass hierarchy case, a superposition of two such features corresponding
to the vacuum neutrino mass-squared differences associated with 
solar and atmospheric neutrino oscillations.
These neutrino spectral features 
offer a unique potential probe of the conditions in the
 supernova environment and may allow us to distinguish between
O-Ne-Mg and Fe core-collapse supernovae.
\end{abstract}
 
\maketitle

In this Letter we suggest an exciting new 
neutrino signal-based probe of 
conditions deep inside a supernova.
We do this by performing the first fully self-coupled
3-neutrino flavor ($3\times3$) evolution calculations.  Stars of
$\sim 8$--$11\,M_\odot$ develop degenerate O-Ne-Mg cores,
at least some of which eventually collapse to produce supernovae
(e.g., Refs.~\cite{Nomoto:1984aa,Nomoto:1987aa,Ritossa:1999aa}). 
The matter density falls off so steeply in the region
between such a core and the hydrogen envelope that there is little
hindrance to the outgoing supernova shock. Consequently,
O-Ne-Mg core-collapse supernovae are the only case where
neutrino-driven explosion has been demonstrated by several
groups \cite{Mayle:1988aa,Kitaura:2005bt,Dessart:2006gd}.
Such supernovae may be the site for producing
the heaviest elements by rapid neutron capture \cite{Ning:2007tu} and may
also explain the observed subluminous supernovae \cite{Kitaura:2005bt}. 
They are expected to be relatively common because the known
progenitors of most core-collapse supernovae lie in the mass range
$\sim 8$--$20\,M_\odot$ (e.g., Ref.~\cite{Li:2005ha}).

The region of steeply-falling
matter density immediately above an O-Ne-Mg core provides an
extremely interesting environment for studying neutrino flavor
evolution. For the vacuum neutrino mass-squared differences
$\dmatm$ and $\dmsol$ associated
with atmospheric and solar neutrino oscillations, respectively,
the two corresponding conventional Mikheyev-Smirnov-Wolfenstein (MSW) 
\cite{Wolfenstein:1977ue,Mikheyev:1985aa} resonances occur with
very small radial separation in this region. As the neutrino number
density decreases much more gently with radius than the matter
density, neutrino self-coupling can affect flavor evolution associated
with both $\dmatm$ and $\dmsol$, and
a full treatment of $3\times3$ 
mixing appears to be required. To identify
clearly any new physics, we study the relatively simple case of the
neutronization burst, which consists of predominantly $\nu_e$
emitted when the shock breaks through the neutrino sphere.

Traditional analyses of flavor evolution of supernova neutrinos
are based on the pure matter-driven MSW effect (see, e.g., 
Refs.~\cite{Kuo:1987qu,Dighe:1999bi,Lunardini:2003eh,Kneller:2005hf}).
The evolution of neutrino flavor state  $|\psi\rangle$ in matter
is described by the Schr\"{o}dinger-like equation,
\begin{equation}
\rmi\frac{\rmd}{\rmd t}|\psi\rangle
=\hat{H}|\psi\rangle,
\end{equation}
where $t$ is an Affine parameter along the neutrino
worldline, and the Hamiltonian $\hat{H}$
is composed of two pieces: $\hat{H}=\hat{H}_\mathrm{vac}+\hat{H}_\mathrm{matt}$.
The matter contribution is
$\langle\nu_\alpha|\hat{H}_\mathrm{matt}|\nu_\beta\rangle=
\sqrt{2}\GF n_e\delta_{\alpha\beta}\delta_{e\alpha}$, where $\GF$ is the Fermi
constant, $n_e$ is the electron number 
density, and $|\nu_\alpha\rangle$ denotes
a pure flavor state with $\alpha=e,\mu,\tau$.
The vacuum piece of $\hat{H}$ is
$\langle\nu_\alpha|\hat{H}_\mathrm{vac}|\nu_\beta\rangle=
(2E_\nu)^{-1}(U M U^\dagger)_{\alpha\beta}$,  where $E_\nu$
is the neutrino energy. The transformation
$U_{\alpha i}$ relates pure flavor state $|\nu_\alpha\rangle$ 
to vacuum mass eigenstate $|\nu_i\rangle$ (see Chap.~13 of Ref.~\cite{PDBook}
for our convention): 
$|\nu_\alpha\rangle=\sum_{i=1,2,3}U^*_{\alpha i}|\nu_i\rangle$.
The mass matrix is diagonal in the vacuum mass basis,
 $M=\mathrm{diag}(0,\Delta m_{21}^2,\Delta m_{21}^2+\Delta m_{32}^2)$,
where the mass-squared differences are
$\Delta m_{ij}^2=m_i^2-m_j^2$.
In calculations presented here
 we take the three mixing angles and the CP violating phase to be
$\theta_{12}=0.6$, $\theta_{23}=\pi/4$, 
$\theta_{13}=0.1$, $\delta=0$, respectively. We take 
$\Delta m_{21}^2=8\times10^{-5}\,\mathrm{eV}^2\simeq\dmsol$ and
$\Delta m_{32}^2=\pm3\times10^{-3}\,\mathrm{eV}^2\simeq\pm\dmatm$,
where the plus (minus) sign is for
the normal (inverted) mass hierarchy.

In pure matter-driven MSW evolution, for small $\theta_{13}$, 
the $\nu_e$ survival probability
$P_{\nu_e\nu_e}=|\langle\nu_e|\psi\rangle|^2$ can be factorized \cite{Kuo:1987qu}:
$P_{\nu_e\nu_e}=P_{\nu_e\nu_e}^\mathrm{H}P_{\nu_e\nu_e}^\mathrm{L}$,
where $P_{\nu_e\nu_e}^\mathrm{H}$ and $P_{\nu_e\nu_e}^\mathrm{L}$
are the $\nu_e$ survival probabilities in 2-flavor ($2\times2$) mixing
processes at the $\dmatm$ and $\dmsol$
scales, respectively.
In other words, the full $3\times3$ MSW result is the superposition
of two independent $2\times2$ MSW scenarios, 
one for each of the solar and atmospheric
mass-squared differences.

\begin{figure}
\includegraphics*[width=\myfigwid, keepaspectratio]{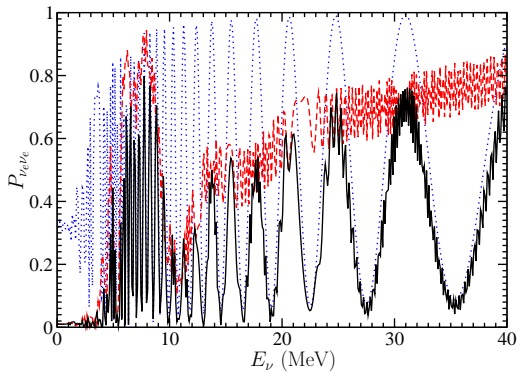} 
\caption{\label{fig:MSW}(Color online)
Neutrino survival probabilities $P_{\nu_e\nu_e}$ as functions
of neutrino energy $E_\nu$ for pure matter-driven MSW evolution.
The $2\times2$ flavor mixing cases with 
$\Delta m^2\simeq\Delta m^2_\mathrm{atm}$ and 
$\Delta m^2_\odot$ are shown as the dashed and dotted lines, respectively.
The $2\times2$ flavor mixing
case with $\Delta m^2\simeq-\Delta m^2_\mathrm{atm}$  (not shown)
corresponds to $P_{\nu_e\nu_e}\simeq1$ for all energies.
The solid line gives $P_{\nu_e\nu_e}(E_\nu)$ for full $3\times3$
flavor mixing with the normal mass hierarchy. The $3\times3$
inverted mass hierarchy case (not shown) is almost
identical to the dotted line.}
\end{figure}

Using the $n_e$ profile for the O-Ne-Mg core model of 
Refs.~\cite{Nomoto:1984aa,Nomoto:1987aa} and the neutrino mixing
parameters given above, we show $P_{\nu_e\nu_e}$ as a function of
$E_\nu$ in Fig.~\ref{fig:MSW} assuming pure matter-driven MSW evolution. 
The results shown are
for radius $r=5000$ km, where the vacuum Hamiltonian dominates
for most neutrino energies. The dashed and dotted lines in this figure
show the $2\times2$ flavor mixing cases with the normal mass hierarchy
for $\Delta m^2=3\times10^{-3}\,\mathrm{eV}^2$ 
($\simeq\dmatm$) and $8\times10^{-5}\,\mathrm{eV}^2$ 
($\simeq\dmsol$), respectively.
In these cases we take the effective $2\times2$ vacuum mixing angles to be
$\theta=0.1$ and $0.6$, respectively.
We note that in either case
MSW flavor transformation for neutronization burst neutrinos of
average energy  $\langle E_{\nu_e}\rangle=11$~MeV
is neither fully adiabatic ($P_{\nu_e\nu_e}=\sin^2\theta$)
nor fully non-adiabatic ($P_{\nu_e\nu_e}=\cos^2\theta$) due to
the rapid decrease of matter density $\rho$ with radius
in the region
of  interest ($|\rmd(\mathrm{ln}\rho)/\rmd r| \gtrsim 0.04\,\mathrm{km}^{-1}$).
The spike in $P_{\nu_e\nu_e}^\mathrm{H}(E_\nu)$
(dashed line) at $E_\nu\simeq8$ MeV is caused by a 
sharp change in $n_e$ at the base of
the hydrogen envelope, where the electron fraction $Y_e$
jumps from $0.5$ to $\sim0.85$.
The $2\times2$ inverted mass hierarchy
case with $\Delta m^2\simeq-\dmatm$
has $P_{\nu_e\nu_e}\simeq1$ for all energies (i.e., no MSW resonance).
In the complete $3\times3$ mixing case with the normal
mass hierarchy, $P_{\nu_e\nu_e}$ is given by the solid line.
This case corresponds closely to a succession of two
independent $2\times2$ mixing schemes, with the solid line
being approximately the product of the values of the
dashed ($P_{\nu_e\nu_e}^\mathrm{H}$) and dotted ($P_{\nu_e\nu_e}^\mathrm{L}$) 
lines. The $3\times 3$ inverted mass hierarchy case gives $P_{\nu_e\nu_e}$
nearly identical to the dotted line.

In supernovae, where  neutrino luminosities are large,
neutrino-neutrino forward scattering contributes another term for
the Hamiltonian \cite{Fuller:1987aa,Pantaleone:1992xh,Sigl:1992fn}
\begin{equation}
\hat{H}_{\nu\nu}=\sqrt{2}\GF\sum_\lambda n_{\nu,\lambda} 
|\psi_\lambda\rangle\langle\psi_\lambda|,
\end{equation}
where $\sum_\lambda$ sums over all background neutrino states 
$|\psi_\lambda\rangle$
with number density $n_{\nu,\lambda}$. 
%
To simplify the problem,
we adopt the ``single-angle approximation'' in which neutrinos emitted in
all directions from the neutrino sphere have the same flavor
evolution histories as those with the same energies but
propagating along a radial trajectory. With this approximation we have
\begin{equation}
\sum_\lambda n_{\nu,\lambda}\longrightarrow
\frac{D(r/R_\nu)}{2\pi R_\nu^2}\frac{L_{\nu_e}}{\langle E_{\nu_e}\rangle}
\int\rmd E_\nu f_{\nu_e}(E_\nu),
\end{equation}
where $D(\xi)=\frac{1}{2}(1-\sqrt{1-\xi^{-2}})^2$.
In our calculations for the neutronization burst
we assume $\nu_e$ is the only neutrino species
emitted from the neutrino sphere (at radius $R_\nu=60$ km) and
take the $\nu_e$ luminosity to be $L_{\nu_e}=10^{53}$ erg/s.
The $\nu_e$ energy distribution function $f_{\nu_e}(E_\nu)$
is taken to be of Fermi-Dirac form with  degeneracy parameter
$\eta=3$ and with an average $\nu_e$ energy
$\langle E_{\nu_e}\rangle=11$ MeV.
Full $2\times2$ multi-angle simulations show that
the single-angle approximation appears to be adequate for
qualitative studies of the collective flavor transformation
phenomena of interest here 
\cite{Duan:2006an,EstebanPretel:2007ec,Fogli:2007bk}.

\begin{figure*}
\begin{center}
$\begin{array}{@{}c@{\hspace{\myfigsep}}c@{}}
\includegraphics*[width=\myfigwid, keepaspectratio]{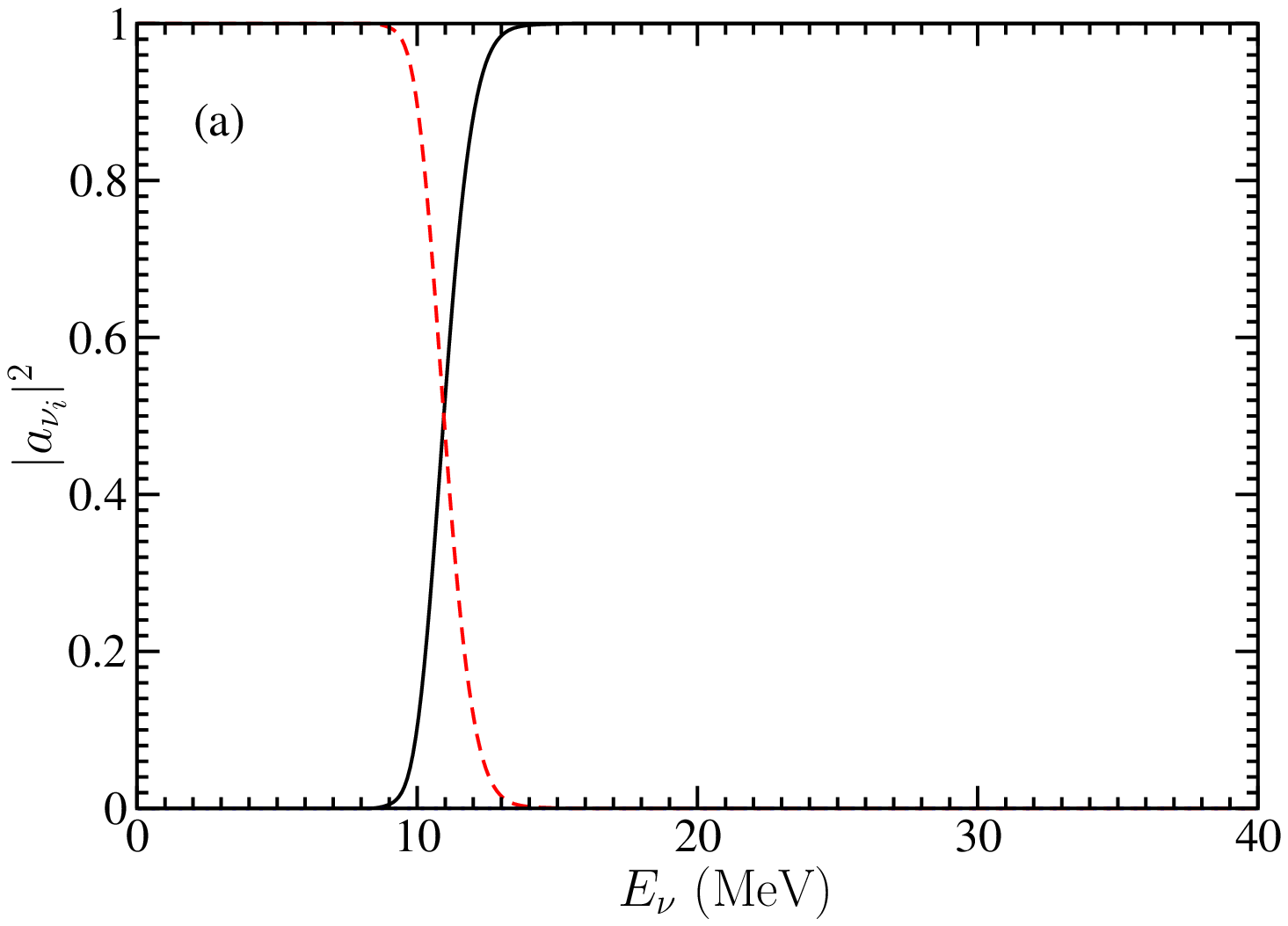} &
\includegraphics*[width=\myfigwid, keepaspectratio]{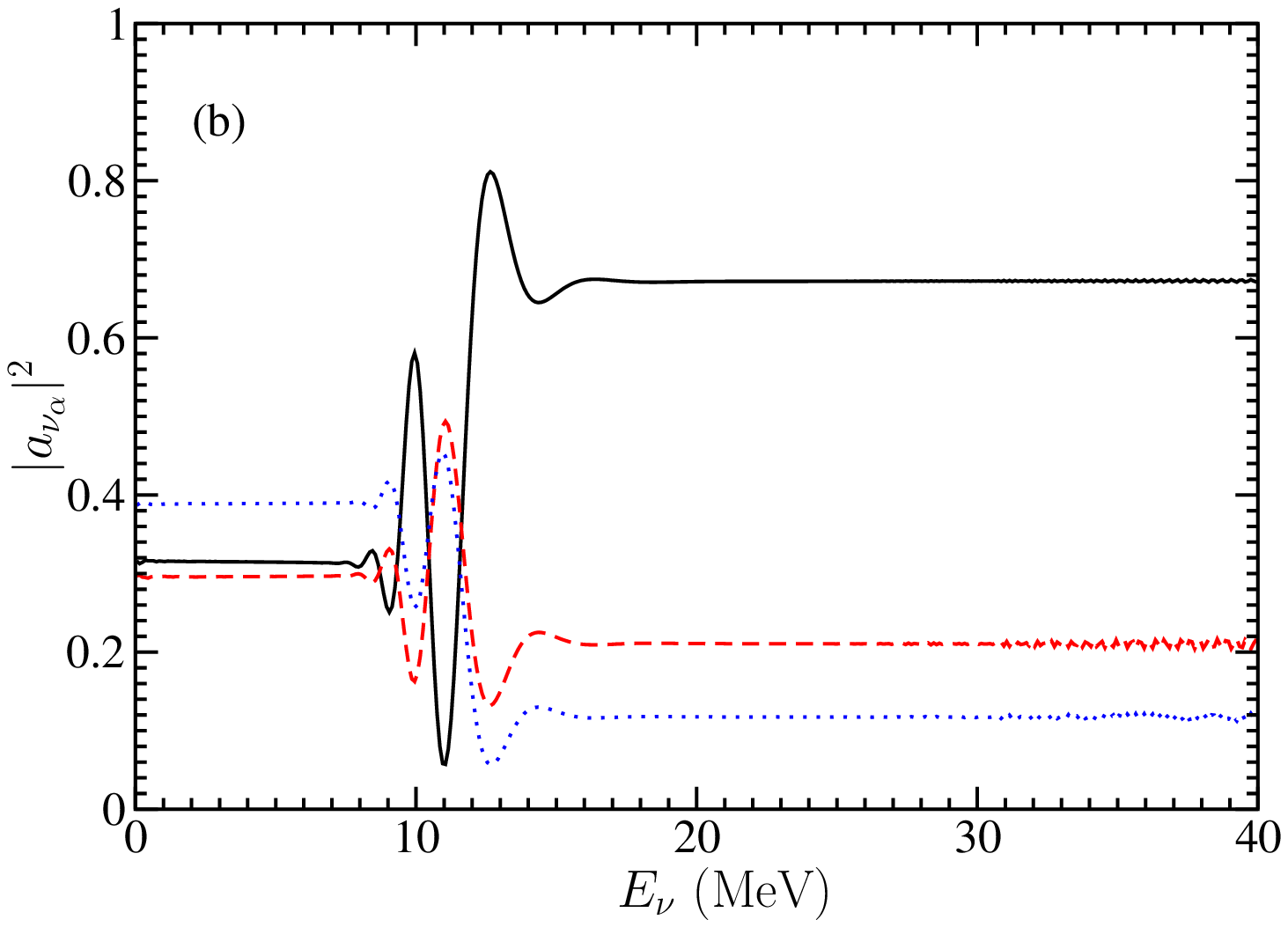} \\
\includegraphics*[width=\myfigwid, keepaspectratio]{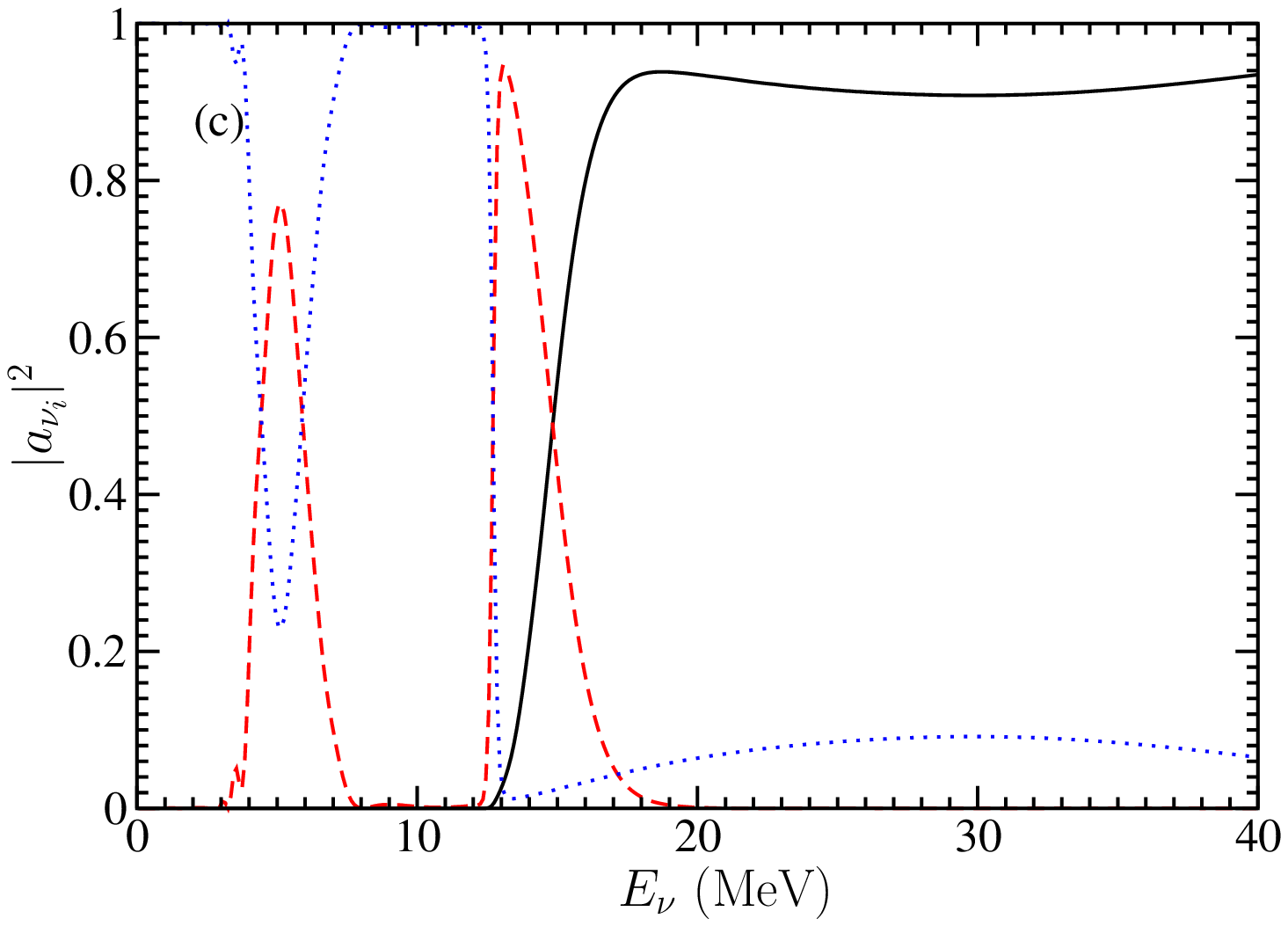} &
\includegraphics*[width=\myfigwid, keepaspectratio]{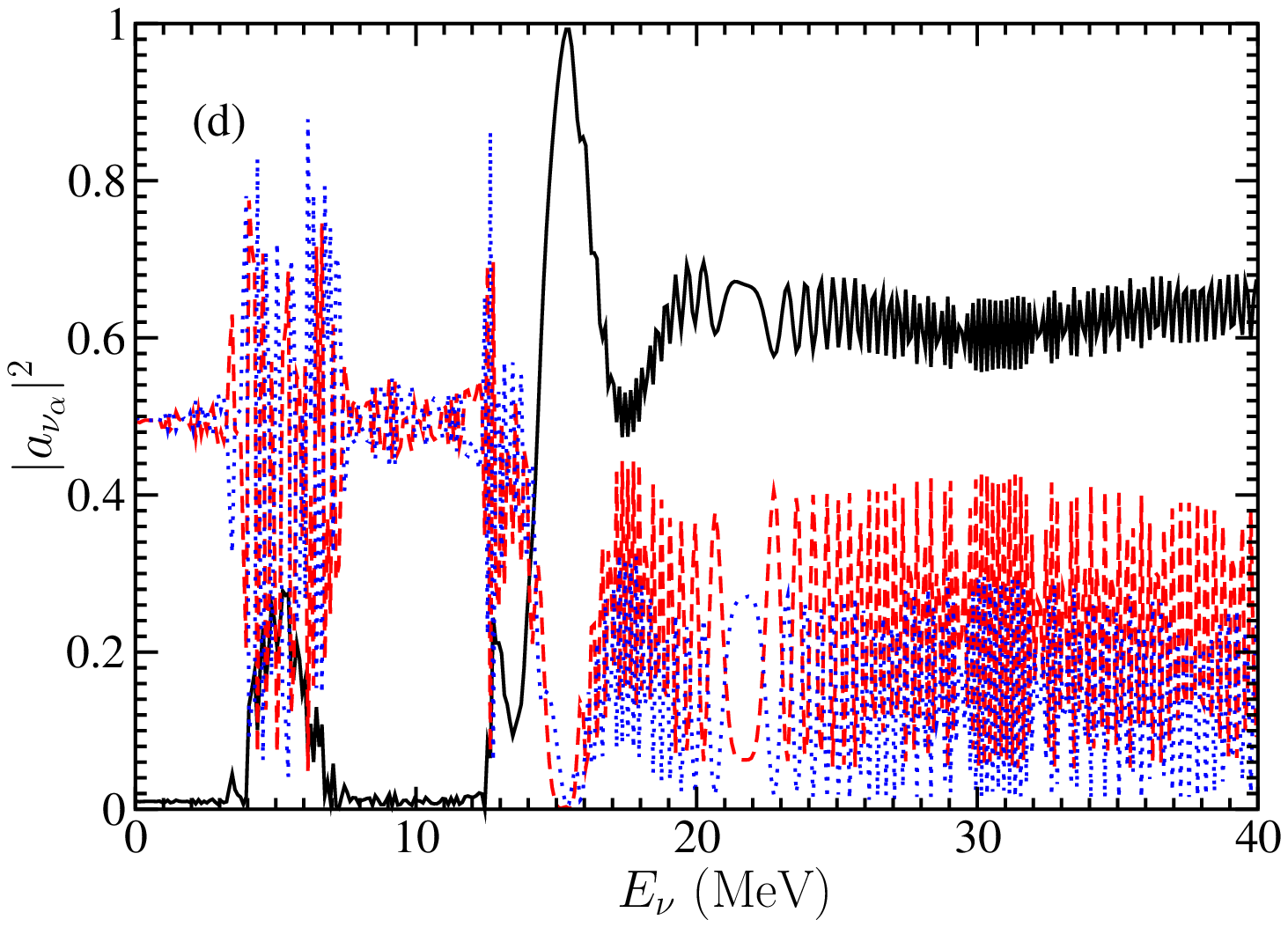}
\end{array}$
\end{center}
\caption{\label{fig:nu-bg}(Color online) 
Probabilities as functions of neutrino energy $E_\nu$ 
for neutrinos to be in each 
vacuum mass eigenstate ($|a_{\nu_i}|^2$, left panels) and flavor eigenstate 
($|a_{\nu_\alpha}|^2$, right panels), respectively. 
The solid, dashed
and dotted lines represent the $\nu_1$, $\nu_2$, and $\nu_3$
states in the left panels and the $\nu_e$, $\nu_\mu$, and $\nu_\tau$
states in the right panels.
Top (bottom) panels show the inverted (normal)  mass hierarchy case.}
\end{figure*}

Fig.~\ref{fig:nu-bg} shows the results of single-angle
simulations of full $3\times3$ neutrino flavor
evolution including nonlinear neutrino self-coupling
for the neutrino mixing and emission
parameters given above. Results for both the
inverted (upper panels) and normal (lower panels)
neutrino mass hierarchies are presented, again at radius $r=5000$ km
as in Fig.~\ref{fig:MSW}. The left-hand panels
show the probability $|a_{\nu_i}|^2=|\langle\nu_i|\psi\rangle|^2$
for neutrinos to be
in  each of the mass eigenstates $|\nu_i\rangle$, and the right-hand
panels show the probability 
$|a_{\nu_\alpha}|^2=|\langle\nu_\alpha|\psi\rangle|^2$
for neutrinos to be in each of the flavor states $|\nu_\alpha\rangle$.

The inverted neutrino mass hierarchy produces a stepwise $\nu_2/\nu_1$
conversion at energy $E_\nu\simeq11$ MeV [Fig.~\ref{fig:nu-bg}(a)].
This spectral swap feature can be understood in
 a $2\times2$ mixing scheme with $\Delta m^2\simeq\dmsol$
(see, e.g., Ref.~\cite{Duan:2007fw}).
In this scheme the flavor evolution of a neutrino
can be represented as the precession of a spin or polarization vector 
in flavor isospace, in analogy to a magnetic spin
(e.g., Ref.~\cite{Sigl:1992fn}). When neutrino number fluxes are large,
the neutrino self-coupling is strong and
the ``magnetic spins'' representing neutrinos can rotate collectively
in the region where a neutrino with a representative energy
would experience a resonance in the pure matter-drive MSW
evolution \cite{Pastor:2002we}.
This corresponds to a
neutrino-background-enhanced MSW-like flavor transformation
\cite{Qian:1994wh,Pastor:2002we,Fuller:2005ae,Duan:2007fw}.
Subsequently, the ``magnetic spins'' will enter a collective
precession mode. As neutrino fluxes become small at large radii and
the collective precession mode dies out, a mass-basis spectral swap
is established \cite{Duan:2007mv,Raffelt:2007cb}. The swap point
in the neutrino energy spectrum is determined by conservation of
a mass-basis ``lepton number'' \cite{Hannestad:2006nj,Raffelt:2007cb}.
In fact, the result of the full $3\times3$ calculation agrees very well 
with that of the $2\times2$ calculation with $\Delta m^2\simeq\dmsol$.
In contrast to the pure-matter driven MSW evolution,
neutrinos on the two sides of the swap point appear to have
experienced almost fully adiabatic or fully non-adiabatic
flavor transformation with those in the low (high) mass eigenstate
ending up on the right (left) hand side of the swap point.
We note that no neutrino-background-enhanced MSW-like flavor transformation
occurs at the $\dmatm$ scale in the inverted mass hierarchy
case. This is analogous to the pure matter-driven MSW evolution. 
We also note that
conservation of the mass-basis lepton number prohibits the formation
of a spectral swap in the corresponding $2\times2$ mixing scheme with 
$\Delta m^2\simeq-\dmatm$
because all neutrinos start as $\nu_e$ in our calculation 
\footnote{A.~Mirizzi (2007), private communication.}.

Because $\theta_{12}$ is large, $P_{\nu_e\nu_e}(E_\nu)$ exhibit
a large oscillatory feature in the transition region near the stepwise
$\nu_2/\nu_1$ conversion point at $E_\nu\sim11$ MeV 
[see Fig.~\ref{fig:nu-bg}(b)]. Outside this region, spectral swap
can also be seen for the neutrino flavor states.
For example, 
$P_{\nu_e\nu_e}\simeq|U_{e2}|^2\simeq0.32$
($|U_{e1}|^2\simeq0.67$)
for $E_\nu\lesssim9$ MeV
($E_\nu\gtrsim16$ MeV).
 Note that $|a_{\nu_\mu}|^2$ is
larger (smaller) than $|a_{\nu_\tau}|^2$ in 
the energy regime $E_\nu\gtrsim13$ MeV
($E_\nu\lesssim10$ MeV). This is a consequence of setting the CP-violating
phase to $\delta=0$. As $\delta$ is increased, $|a_{\nu_\tau}|^2$ 
increases (decreases) for $E_\nu\gtrsim13$ MeV
($E_\nu\lesssim10$ MeV), and $|a_{\nu_\mu}|^2=|a_{\nu_\tau}|^2$
for $\delta=\pi/2$. For $\delta=\pi$, the $|a_{\nu_\mu}|^2$ and
$|a_{\nu_\tau}|^2$ curves in Fig.~\ref{fig:nu-bg}(b) switch places.

Fig.~\ref{fig:nu-bg}(c) shows that the normal neutrino mass hierarchy
produces a superposition of two spectral swap features,
reminiscent of the factorization property of pure
matter-driven MSW evolution. The $\nu_3/\nu_2$ swap at
$E_\nu\simeq12.7$ MeV and the $\nu_2/\nu_1$ swap at
$E_\nu\simeq15$ MeV correspond to those in
the $2\times2$ schemes with $\Delta m^2\simeq \dmatm$
and $\dmsol$, respectively.
This result seems to justify the $2\times2$
approximation used in previous work 
(e.g., Refs.\cite{Pastor:2002we,Balantekin:2004ug,%
Fuller:2005ae,Duan:2005cp,Duan:2006an,Duan:2006jv,%
EstebanPretel:2007ec,Duan:2007bt,Fogli:2007bk}),
but is somewhat surprising
given the nonlinear nature of neutrino self-coupling
and the fact that the regions of collective flavor transformation
for $\dmatm$ and $\dmsol$ overlap with each other.
We note that the $\nu_3/\nu_2$ swap is much sharper than the 
$\nu_2/\nu_1$ swap. We also note that
the dip (bump) in $|a_{\nu_3}|^2$ ($|a_{\nu_2}|^2$) 
centered at $E_\nu\simeq5.2$ MeV
corresponds to the abrupt change in $n_e$ at the base of the hydrogen
envelope. This feature in the $n_e$ profile reduces the efficiency of the 
neutrino-background-enhanced MSW-like transformation of $\nu_e$ 
at the $\dmatm$ scale. A similar feature is also
present in the pure matter-driven MSW evolution, but
in a different energy range (see Fig.~\ref{fig:MSW}).

As in the inverted mass hierarchy case,  the spectral swaps
are also present in the flavor basis
[see Fig.~\ref{fig:nu-bg}(d)].
Except for the moderate bump centered at $E_\nu\simeq5.2$ MeV, nearly
all $\nu_e$'s are transformed 
(with $P_{\nu_e\nu_e}\simeq|U_{e3}|^2\simeq0.01$) below the
$\nu_3/\nu_2$ swap energy $E_\nu\simeq12.7$ MeV.
In contrast, there is less significant $\nu_e$ depletion above 
the $\nu_2/\nu_1$ swap energy $E_\nu\simeq15$ MeV, 
for which energy range
$P_{\nu_e\nu_e}\simeq|U_{e1}|^2\simeq0.67$ would be expected.
The bump in $P_{\nu_e\nu_e}(E_\nu)$ at $E_\nu\simeq5.2$ MeV corresponds to 
the feature in $|a_{\nu_2}|^2$ at the same energy as explained above.
On the other hand, we note that the apparent peak of $P_{\nu_e\nu_e}(E_\nu)$ 
in $E_\nu\simeq15.5\,\mathrm{MeV}$ is part of the same oscillation feature
discussed above for Fig.~\ref{fig:nu-bg}(b).
This oscillation feature will disappear at very large distances
where coherence has been lost.

The spectral swap features illustrated here for the neutronization burst
from an O-Ne-Mg core-collapse supernova are
not expected to be present for an Fe-core collapse supernova.
This is because at the neutronization burst epoch,
there is an extended region of high $n_e$ above an Fe core.
So in this case high neutrino fluxes are always accompanied by high $n_e$,
which inhibits neutrino-background-enhanced MSW-like flavor transformation.
The studies of neutrinos from Fe core-collapse supernovae
based on pure matter-driven MSW evolution show that
the $\nu_e$ survival probability is
$P_{\nu_e\nu_e}\simeq\sin^2\theta_\odot\simeq0.32$ or less
for either mass hierarchy 
(see, e.g., Table I in Ref.~\cite{Kachelriess:2004ds}).
If enough high-energy ($E_\nu\gtrsim15$ MeV) neutrino events 
are collected from the neutronization burst of a future Galactic supernova
in both the charged-current and neutral-current channels,
the progenitor may be identified as having an O-Ne-Mg or Fe core
based on whether or not $\nu_e$ is the dominant species
in the burst.  Because the total neutrino fluence
over the $\sim10$ ms duration of the neutronization burst
is only a small fraction of that emitted during 
the first several seconds after the onset of core collapse, collection of
the required number of events from the neutronization
burst may be beyond the capabilities of existing
detectors but could be within the reach of proposed
megaton water Cherenkov detectors \cite{Itow:2001ee,Goodman:2001aq,Kachelriess:2004ds}
and liquid argon detectors like ICARUS \cite{GilBotella:2003sz,Vignoli:2006jz}.
The low-energy ($\lesssim10$ MeV) neutrino signals
in the O-Ne-Mg core-collapse supernova neutronization burst,
though even more difficult to detect,
carry information that potentially can distinguish between
the neutrino mass hierarchies [see Figs.~\ref{fig:nu-bg}(b) and (d)].



\begin{acknowledgments}
We thank K.~Nomoto for providing the electron number density
profile in his O-Ne-Mg core model.
We appreciate the hospitality of the Institute for Nuclear Theory
at the University of Washington
during the completion of this work.
This work was supported in part by 
NSF grant PHY-04-00359 at UCSD,
DOE grants DE-FG02-87ER40328 at UMN,
DE-FG02-00ER41132 at INT,
an IGPP/LANL mini-grant,
and by the DOE Office of Nuclear Physics, the LDRD Program
and Open Supercomputing at LANL.
\end{acknowledgments}

\bibliography{ref}

\begin{thebibliography}{37}
\expandafter\ifx\csname natexlab\endcsname\relax\def\natexlab#1{#1}\fi
\expandafter\ifx\csname bibnamefont\endcsname\relax
  \def\bibnamefont#1{#1}\fi
\expandafter\ifx\csname bibfnamefont\endcsname\relax
  \def\bibfnamefont#1{#1}\fi
\expandafter\ifx\csname citenamefont\endcsname\relax
  \def\citenamefont#1{#1}\fi
\expandafter\ifx\csname url\endcsname\relax
  \def\url#1{\texttt{#1}}\fi
\expandafter\ifx\csname urlprefix\endcsname\relax\def\urlprefix{URL }\fi
\providecommand{\bibinfo}[2]{#2}
\providecommand{\eprint}[2][]{\url{#2}}

\bibitem[{\citenamefont{Nomoto}(1984)}]{Nomoto:1984aa}
\bibinfo{author}{\bibfnamefont{K.}~\bibnamefont{Nomoto}},
  \bibinfo{journal}{Astrophys. J.} \textbf{\bibinfo{volume}{277}},
  \bibinfo{pages}{791} (\bibinfo{year}{1984}).

\bibitem[{\citenamefont{Nomoto}(1987)}]{Nomoto:1987aa}
\bibinfo{author}{\bibfnamefont{K.}~\bibnamefont{Nomoto}},
  \bibinfo{journal}{Astrophys. J.} \textbf{\bibinfo{volume}{322}},
  \bibinfo{pages}{206} (\bibinfo{year}{1987}).

\bibitem[{\citenamefont{{Ritossa} et~al.}(1999)\citenamefont{{Ritossa},
  {Garc{\'{\i}}a-Berro}, and {Iben}}}]{Ritossa:1999aa}
\bibinfo{author}{\bibfnamefont{C.}~\bibnamefont{{Ritossa}}},
  \bibinfo{author}{\bibfnamefont{E.}~\bibnamefont{{Garc{\'{\i}}a-Berro}}},
  \bibnamefont{and} \bibinfo{author}{\bibfnamefont{I.~J.}
  \bibnamefont{{Iben}}}, \bibinfo{journal}{Astrophys. J.}
  \textbf{\bibinfo{volume}{515}}, \bibinfo{pages}{381} (\bibinfo{year}{1999}).

\bibitem[{\citenamefont{{Mayle} and {Wilson}}(1988)}]{Mayle:1988aa}
\bibinfo{author}{\bibfnamefont{R.}~\bibnamefont{{Mayle}}} \bibnamefont{and}
  \bibinfo{author}{\bibfnamefont{J.~R.} \bibnamefont{{Wilson}}},
  \bibinfo{journal}{Astrophys. J.} \textbf{\bibinfo{volume}{334}},
  \bibinfo{pages}{909} (\bibinfo{year}{1988}).

\bibitem[{\citenamefont{Kitaura et~al.}(2006)\citenamefont{Kitaura, Janka, and
  Hillebrandt}}]{Kitaura:2005bt}
\bibinfo{author}{\bibfnamefont{F.~S.} \bibnamefont{Kitaura}},
  \bibinfo{author}{\bibfnamefont{H.-T.} \bibnamefont{Janka}}, \bibnamefont{and}
  \bibinfo{author}{\bibfnamefont{W.}~\bibnamefont{Hillebrandt}},
  \bibinfo{journal}{Astron. Astrophys.} \textbf{\bibinfo{volume}{450}},
  \bibinfo{pages}{345} (\bibinfo{year}{2006}), \eprint{astro-ph/0512065}.

\bibitem[{\citenamefont{Dessart et~al.}(2006)}]{Dessart:2006gd}
\bibinfo{author}{\bibfnamefont{L.}~\bibnamefont{Dessart}} \bibnamefont{et~al.},
  \bibinfo{journal}{Astrophys. J.} \textbf{\bibinfo{volume}{644}},
  \bibinfo{pages}{1063} (\bibinfo{year}{2006}), \eprint{astro-ph/0601603}.

\bibitem[{\citenamefont{Ning et~al.}(2007)\citenamefont{Ning, Qian, and
  Meyer}}]{Ning:2007tu}
\bibinfo{author}{\bibfnamefont{H.}~\bibnamefont{Ning}},
  \bibinfo{author}{\bibfnamefont{Y.~Z.} \bibnamefont{Qian}}, \bibnamefont{and}
  \bibinfo{author}{\bibfnamefont{B.~S.} \bibnamefont{Meyer}},
  \bibinfo{journal}{Astrophys. J.} \textbf{\bibinfo{volume}{667}},
  \bibinfo{pages}{L159} (\bibinfo{year}{2007}), \eprint{arXiv:0708.1748
  [astro-ph]}.

\bibitem[{\citenamefont{Li et~al.}(2006)}]{Li:2005ha}
\bibinfo{author}{\bibfnamefont{W.-D.} \bibnamefont{Li}} \bibnamefont{et~al.},
  \bibinfo{journal}{Astrophys. J.} \textbf{\bibinfo{volume}{641}},
  \bibinfo{pages}{1060} (\bibinfo{year}{2006}), \eprint{astro-ph/0507394}.

\bibitem[{\citenamefont{Wolfenstein}(1978)}]{Wolfenstein:1977ue}
\bibinfo{author}{\bibfnamefont{L.}~\bibnamefont{Wolfenstein}},
  \bibinfo{journal}{Phys. Rev.} \textbf{\bibinfo{volume}{D17}},
  \bibinfo{pages}{2369} (\bibinfo{year}{1978}).

\bibitem[{\citenamefont{Mikheyev and Smirnov}(1985)}]{Mikheyev:1985aa}
\bibinfo{author}{\bibfnamefont{S.~P.} \bibnamefont{Mikheyev}} \bibnamefont{and}
  \bibinfo{author}{\bibfnamefont{A.~Y.} \bibnamefont{Smirnov}},
  \bibinfo{journal}{Yad. Fiz.} \textbf{\bibinfo{volume}{42}},
  \bibinfo{pages}{1441} (\bibinfo{year}{1985}), \bibinfo{note}{[Sov. J. Nucl.
  Phys. 42, 913 (1985)]}.

\bibitem[{\citenamefont{Kuo and Pantaleone}(1988)}]{Kuo:1987qu}
\bibinfo{author}{\bibfnamefont{T.-K.} \bibnamefont{Kuo}} \bibnamefont{and}
  \bibinfo{author}{\bibfnamefont{J.~T.} \bibnamefont{Pantaleone}},
  \bibinfo{journal}{Phys. Rev.} \textbf{\bibinfo{volume}{D37}},
  \bibinfo{pages}{298} (\bibinfo{year}{1988}).

\bibitem[{\citenamefont{Dighe and Smirnov}(2000)}]{Dighe:1999bi}
\bibinfo{author}{\bibfnamefont{A.~S.} \bibnamefont{Dighe}} \bibnamefont{and}
  \bibinfo{author}{\bibfnamefont{A.~Y.} \bibnamefont{Smirnov}},
  \bibinfo{journal}{Phys. Rev.} \textbf{\bibinfo{volume}{D62}},
  \bibinfo{pages}{033007} (\bibinfo{year}{2000}), \eprint{hep-ph/9907423}.

\bibitem[{\citenamefont{Lunardini and Smirnov}(2003)}]{Lunardini:2003eh}
\bibinfo{author}{\bibfnamefont{C.}~\bibnamefont{Lunardini}} \bibnamefont{and}
  \bibinfo{author}{\bibfnamefont{A.~Y.} \bibnamefont{Smirnov}},
  \bibinfo{journal}{JCAP} \textbf{\bibinfo{volume}{0306}}, \bibinfo{pages}{009}
  (\bibinfo{year}{2003}), \eprint{hep-ph/0302033}.

\bibitem[{\citenamefont{Kneller and McLaughlin}(2006)}]{Kneller:2005hf}
\bibinfo{author}{\bibfnamefont{J.~P.} \bibnamefont{Kneller}} \bibnamefont{and}
  \bibinfo{author}{\bibfnamefont{G.~C.} \bibnamefont{McLaughlin}},
  \bibinfo{journal}{Phys. Rev.} \textbf{\bibinfo{volume}{D73}},
  \bibinfo{pages}{056003} (\bibinfo{year}{2006}), \eprint{hep-ph/0509356}.

\bibitem[{\citenamefont{Yao et~al.}(2006)}]{PDBook}
\bibinfo{author}{\bibfnamefont{W.-M.} \bibnamefont{Yao}} \bibnamefont{et~al.},
  \bibinfo{journal}{J.~Phys.} \textbf{\bibinfo{volume}{G33}},
  \bibinfo{pages}{1} (\bibinfo{year}{2006}).

\bibitem[{\citenamefont{Fuller et~al.}(1987)\citenamefont{Fuller, Mayle,
  Wilson, and Schramm}}]{Fuller:1987aa}
\bibinfo{author}{\bibfnamefont{G.~M.} \bibnamefont{Fuller}},
  \bibinfo{author}{\bibfnamefont{R.~W.} \bibnamefont{Mayle}},
  \bibinfo{author}{\bibfnamefont{J.~R.} \bibnamefont{Wilson}},
  \bibnamefont{and} \bibinfo{author}{\bibfnamefont{D.~N.}
  \bibnamefont{Schramm}}, \bibinfo{journal}{Astrophys. J.}
  \textbf{\bibinfo{volume}{322}}, \bibinfo{pages}{795} (\bibinfo{year}{1987}).

\bibitem[{\citenamefont{Pantaleone}(1992)}]{Pantaleone:1992xh}
\bibinfo{author}{\bibfnamefont{J.~T.} \bibnamefont{Pantaleone}},
  \bibinfo{journal}{Phys. Rev.} \textbf{\bibinfo{volume}{D46}},
  \bibinfo{pages}{510} (\bibinfo{year}{1992}).

\bibitem[{\citenamefont{Sigl and Raffelt}(1993)}]{Sigl:1992fn}
\bibinfo{author}{\bibfnamefont{G.}~\bibnamefont{Sigl}} \bibnamefont{and}
  \bibinfo{author}{\bibfnamefont{G.}~\bibnamefont{Raffelt}},
  \bibinfo{journal}{Nucl. Phys.} \textbf{\bibinfo{volume}{B406}},
  \bibinfo{pages}{423} (\bibinfo{year}{1993}).

\bibitem[{\citenamefont{Duan et~al.}(2006{\natexlab{a}})\citenamefont{Duan,
  Fuller, Carlson, and Qian}}]{Duan:2006an}
\bibinfo{author}{\bibfnamefont{H.}~\bibnamefont{Duan}},
  \bibinfo{author}{\bibfnamefont{G.~M.} \bibnamefont{Fuller}},
  \bibinfo{author}{\bibfnamefont{J.}~\bibnamefont{Carlson}}, \bibnamefont{and}
  \bibinfo{author}{\bibfnamefont{Y.-Z.} \bibnamefont{Qian}},
  \bibinfo{journal}{Phys. Rev.} \textbf{\bibinfo{volume}{D74}},
  \bibinfo{pages}{105014} (\bibinfo{year}{2006}{\natexlab{a}}),
  \eprint{astro-ph/0606616}.

\bibitem[{\citenamefont{Esteban-Pretel
  et~al.}(2007)\citenamefont{Esteban-Pretel, Pastor, Tomas, Raffelt, and
  Sigl}}]{EstebanPretel:2007ec}
\bibinfo{author}{\bibfnamefont{A.}~\bibnamefont{Esteban-Pretel}},
  \bibinfo{author}{\bibfnamefont{S.}~\bibnamefont{Pastor}},
  \bibinfo{author}{\bibfnamefont{R.}~\bibnamefont{Tomas}},
  \bibinfo{author}{\bibfnamefont{G.~G.} \bibnamefont{Raffelt}},
  \bibnamefont{and} \bibinfo{author}{\bibfnamefont{G.}~\bibnamefont{Sigl}},
  \bibinfo{journal}{Phys. Rev.} \textbf{\bibinfo{volume}{D76}},
  \bibinfo{pages}{125018} (\bibinfo{year}{2007}), \eprint{arXiv:0706.2498
  [astro-ph]}.

\bibitem[{\citenamefont{Fogli et~al.}(2007)\citenamefont{Fogli, Lisi, Marrone,
  and Mirizzi}}]{Fogli:2007bk}
\bibinfo{author}{\bibfnamefont{G.~L.} \bibnamefont{Fogli}},
  \bibinfo{author}{\bibfnamefont{E.}~\bibnamefont{Lisi}},
  \bibinfo{author}{\bibfnamefont{A.}~\bibnamefont{Marrone}}, \bibnamefont{and}
  \bibinfo{author}{\bibfnamefont{A.}~\bibnamefont{Mirizzi}}
  (\bibinfo{year}{2007}), \eprint{arXiv:0707.1998 [hep-ph]}.

\bibitem[{\citenamefont{Duan et~al.}(2007{\natexlab{a}})\citenamefont{Duan,
  Fuller, and Qian}}]{Duan:2007fw}
\bibinfo{author}{\bibfnamefont{H.}~\bibnamefont{Duan}},
  \bibinfo{author}{\bibfnamefont{G.~M.} \bibnamefont{Fuller}},
  \bibnamefont{and} \bibinfo{author}{\bibfnamefont{Y.-Z.} \bibnamefont{Qian}},
  \bibinfo{journal}{Phys. Rev.} \textbf{\bibinfo{volume}{D76}},
  \bibinfo{pages}{085013} (\bibinfo{year}{2007}{\natexlab{a}}),
  \eprint{arXiv:0706.4293 [astro-ph]}.

\bibitem[{\citenamefont{Pastor and Raffelt}(2002)}]{Pastor:2002we}
\bibinfo{author}{\bibfnamefont{S.}~\bibnamefont{Pastor}} \bibnamefont{and}
  \bibinfo{author}{\bibfnamefont{G.}~\bibnamefont{Raffelt}},
  \bibinfo{journal}{Phys. Rev. Lett.} \textbf{\bibinfo{volume}{89}},
  \bibinfo{pages}{191101} (\bibinfo{year}{2002}), \eprint{astro-ph/0207281}.

\bibitem[{\citenamefont{Qian and Fuller}(1995)}]{Qian:1994wh}
\bibinfo{author}{\bibfnamefont{Y.~Z.} \bibnamefont{Qian}} \bibnamefont{and}
  \bibinfo{author}{\bibfnamefont{G.~M.} \bibnamefont{Fuller}},
  \bibinfo{journal}{Phys. Rev.} \textbf{\bibinfo{volume}{D51}},
  \bibinfo{pages}{1479} (\bibinfo{year}{1995}), \eprint{astro-ph/9406073}.

\bibitem[{\citenamefont{Fuller and Qian}(2006)}]{Fuller:2005ae}
\bibinfo{author}{\bibfnamefont{G.~M.} \bibnamefont{Fuller}} \bibnamefont{and}
  \bibinfo{author}{\bibfnamefont{Y.-Z.} \bibnamefont{Qian}},
  \bibinfo{journal}{Phys. Rev.} \textbf{\bibinfo{volume}{D73}},
  \bibinfo{pages}{023004} (\bibinfo{year}{2006}), \eprint{astro-ph/0505240}.

\bibitem[{\citenamefont{Duan et~al.}(2007{\natexlab{b}})\citenamefont{Duan,
  Fuller, Carlson, and Qian}}]{Duan:2007mv}
\bibinfo{author}{\bibfnamefont{H.}~\bibnamefont{Duan}},
  \bibinfo{author}{\bibfnamefont{G.~M.} \bibnamefont{Fuller}},
  \bibinfo{author}{\bibfnamefont{J.}~\bibnamefont{Carlson}}, \bibnamefont{and}
  \bibinfo{author}{\bibfnamefont{Y.-Z.} \bibnamefont{Qian}},
  \bibinfo{journal}{Phys. Rev.} \textbf{\bibinfo{volume}{D75}},
  \bibinfo{pages}{125005} (\bibinfo{year}{2007}{\natexlab{b}}),
  \eprint{astro-ph/0703776}.

\bibitem[{\citenamefont{Raffelt and Smirnov}(2007)}]{Raffelt:2007cb}
\bibinfo{author}{\bibfnamefont{G.~G.} \bibnamefont{Raffelt}} \bibnamefont{and}
  \bibinfo{author}{\bibfnamefont{A.~Y.} \bibnamefont{Smirnov}},
  \bibinfo{journal}{Phys. Rev.} \textbf{\bibinfo{volume}{D76}},
  \bibinfo{pages}{081301(R)} (\bibinfo{year}{2007}), \eprint{arXiv:0705.1830
  [hep-ph]}.

\bibitem[{\citenamefont{Hannestad et~al.}(2006)\citenamefont{Hannestad,
  Raffelt, Sigl, and Wong}}]{Hannestad:2006nj}
\bibinfo{author}{\bibfnamefont{S.}~\bibnamefont{Hannestad}},
  \bibinfo{author}{\bibfnamefont{G.~G.} \bibnamefont{Raffelt}},
  \bibinfo{author}{\bibfnamefont{G.}~\bibnamefont{Sigl}}, \bibnamefont{and}
  \bibinfo{author}{\bibfnamefont{Y.~Y.~Y.} \bibnamefont{Wong}},
  \bibinfo{journal}{Phys. Rev.} \textbf{\bibinfo{volume}{D74}},
  \bibinfo{pages}{105010} (\bibinfo{year}{2006}), \eprint{astro-ph/0608695}.

\bibitem[{\citenamefont{Balantekin and Y\"{u}ksel}(2005)}]{Balantekin:2004ug}
\bibinfo{author}{\bibfnamefont{A.~B.} \bibnamefont{Balantekin}}
  \bibnamefont{and}
  \bibinfo{author}{\bibfnamefont{H.}~\bibnamefont{Y\"{u}ksel}},
  \bibinfo{journal}{New J. Phys.} \textbf{\bibinfo{volume}{7}},
  \bibinfo{pages}{51} (\bibinfo{year}{2005}), \eprint{astro-ph/0411159}.

\bibitem[{\citenamefont{Duan et~al.}(2006{\natexlab{b}})\citenamefont{Duan,
  Fuller, and Qian}}]{Duan:2005cp}
\bibinfo{author}{\bibfnamefont{H.}~\bibnamefont{Duan}},
  \bibinfo{author}{\bibfnamefont{G.~M.} \bibnamefont{Fuller}},
  \bibnamefont{and} \bibinfo{author}{\bibfnamefont{Y.-Z.} \bibnamefont{Qian}},
  \bibinfo{journal}{Phys. Rev.} \textbf{\bibinfo{volume}{D74}},
  \bibinfo{pages}{123004} (\bibinfo{year}{2006}{\natexlab{b}}),
  \eprint{astro-ph/0511275}.

\bibitem[{\citenamefont{Duan et~al.}(2006{\natexlab{c}})\citenamefont{Duan,
  Fuller, Carlson, and Qian}}]{Duan:2006jv}
\bibinfo{author}{\bibfnamefont{H.}~\bibnamefont{Duan}},
  \bibinfo{author}{\bibfnamefont{G.~M.} \bibnamefont{Fuller}},
  \bibinfo{author}{\bibfnamefont{J.}~\bibnamefont{Carlson}}, \bibnamefont{and}
  \bibinfo{author}{\bibfnamefont{Y.-Z.} \bibnamefont{Qian}},
  \bibinfo{journal}{Phys. Rev. Lett.} \textbf{\bibinfo{volume}{97}},
  \bibinfo{pages}{241101} (\bibinfo{year}{2006}{\natexlab{c}}),
  \eprint{astro-ph/0608050}.

\bibitem[{\citenamefont{Duan et~al.}(2007{\natexlab{c}})\citenamefont{Duan,
  Fuller, Carlson, and Qian}}]{Duan:2007bt}
\bibinfo{author}{\bibfnamefont{H.}~\bibnamefont{Duan}},
  \bibinfo{author}{\bibfnamefont{G.~M.} \bibnamefont{Fuller}},
  \bibinfo{author}{\bibfnamefont{J.}~\bibnamefont{Carlson}}, \bibnamefont{and}
  \bibinfo{author}{\bibfnamefont{Y.-Z.} \bibnamefont{Qian}},
  \bibinfo{journal}{Phys. Rev. Lett.} \textbf{\bibinfo{volume}{99}},
  \bibinfo{pages}{241802} (\bibinfo{year}{2007}{\natexlab{c}}),
  \eprint{arXiv:0707.0290 [astro-ph]}.

\bibitem[{\citenamefont{Kachelriess et~al.}(2005)}]{Kachelriess:2004ds}
\bibinfo{author}{\bibfnamefont{M.}~\bibnamefont{Kachelriess}}
  \bibnamefont{et~al.}, \bibinfo{journal}{Phys. Rev.}
  \textbf{\bibinfo{volume}{D71}}, \bibinfo{pages}{063003}
  (\bibinfo{year}{2005}), \eprint{astro-ph/0412082}.

\bibitem[{\citenamefont{Itow et~al.}(2001)}]{Itow:2001ee}
\bibinfo{author}{\bibfnamefont{Y.}~\bibnamefont{Itow}} \bibnamefont{et~al.}
  (\bibinfo{collaboration}{The T2K}) (\bibinfo{year}{2001}),
  \eprint{hep-ex/0106019}.

\bibitem[{\citenamefont{Goodman et~al.}(2001)}]{Goodman:2001aq}
\bibinfo{author}{\bibfnamefont{M.}~\bibnamefont{Goodman}} \bibnamefont{et~al.},
  \bibinfo{journal}{UNO whitepaper}  (\bibinfo{year}{2001}),
  \eprint{PREPRINT-SBHEP01-3}.

\bibitem[{\citenamefont{Gil-Botella and Rubbia}(2003)}]{GilBotella:2003sz}
\bibinfo{author}{\bibfnamefont{I.}~\bibnamefont{Gil-Botella}} \bibnamefont{and}
  \bibinfo{author}{\bibfnamefont{A.}~\bibnamefont{Rubbia}},
  \bibinfo{journal}{JCAP} \textbf{\bibinfo{volume}{0310}}, \bibinfo{pages}{009}
  (\bibinfo{year}{2003}), \eprint{hep-ph/0307244}.

\bibitem[{\citenamefont{Vignoli et~al.}(2006)\citenamefont{Vignoli, Barni,
  Disdier, Rampoldi, and Passardi}}]{Vignoli:2006jz}
\bibinfo{author}{\bibfnamefont{C.}~\bibnamefont{Vignoli}},
  \bibinfo{author}{\bibfnamefont{D.}~\bibnamefont{Barni}},
  \bibinfo{author}{\bibfnamefont{J.~M.} \bibnamefont{Disdier}},
  \bibinfo{author}{\bibfnamefont{D.}~\bibnamefont{Rampoldi}}, \bibnamefont{and}
  \bibinfo{author}{\bibfnamefont{G.}~\bibnamefont{Passardi}}
  (\bibinfo{collaboration}{ICARUS}), \bibinfo{journal}{AIP Conf. Proc.}
  \textbf{\bibinfo{volume}{823}}, \bibinfo{pages}{1643} (\bibinfo{year}{2006}).

\end{thebibliography}

\end{document}